\documentclass[aps,prl,reprint,groupedaddress,showpacs,amsmath,amssymb,twocolumn,floatfix]{revtex4-1}
\usepackage{graphicx}
\usepackage{bm}
\usepackage{color}
\usepackage[normalem]{ulem}

\begin{document}

\title{High quality magnetotransport in graphene using the edge-free Corbino geometry}

\author{Y. Zeng$^{1}$}
\email{Y.Zeng and J.I.A.Li contributed equally to this work}
\author{J.I.A. Li$^{1}$,$^{*}$}
\author{S.A. Dietrich$^{1}$}
\author{O.M. Ghosh$^{1}$}
\author{K. Watanabe$^{2}$}
\author{T. Taniguchi$^{2}$}
\author{J. Hone$^{3}$}
\author{C.R. Dean$^{1}$}
\email{cory.dean@gmail.com}

\affiliation{$^{1}$Department of Physics, Columbia University, New York, NY, USA}
\affiliation{$^{2}$National Institute for Materials Science, 1-1 Namiki, Tsukuba, Japan}
\affiliation{$^{3}$Department of Mechanical Engineering, Columbia University, New York, NY, USA}

\date{\today}

\begin{abstract}

We report fabrication of graphene devices in a Corbino geometry consisting of concentric circular electrodes with no physical edge connecting the inner and outer electrodes. High device mobility is realized using boron nitride encapsulation together with a dual-graphite gate structure.  Bulk conductance measurement in the quantum Hall effect (QHE) regime outperforms previously reported Hall bar measurements, with improved resolution observed for both the integer and fractional QHE states. We identify apparent phase transitions in the fractional sequence in both the lowest and first excited Landau levels (LLs) and observed features consistent with electron solid phases in higher LLs. 
\end{abstract}

\maketitle

The quantum Hall effect (QHE), characterized by vanishing longitudinal resistance simultaneous with quantized transverse Hall resistance ~\cite{Klitzing1980,Cui_FQHE}, represents one of the most robust examples of 2D topological phenomenon in which an insulating bulk state with non-trivial topological order is separated from the surrounding vacuum by conducting edge modes ~\cite{TKNN1982}. The edge modes associated with the QHE are chiral and therefore dissipationless at all length scales.  Moreover, the transverse Hall resistance, quantized in units of $h/e^{2}$, provides a direct measure of the topological order and is insensitive to details of the sample geometry. In samples with very low disorder, new correlated phases, resulting from strong electron interactions, can be observed outside of the IQHE sequence.  These include the fractional quantum Hall effect (FQHE) liquid states ~\cite{Cui_FQHE,Laughlin1983}, appearing at fractional Landau filling, and with fractionally valued Hall resistance plateaus, and interaction-driven electron solid phases, appearing at fractional filling but with re-entrant integer valued Hall quantization ~\cite{Eisenstein2002bubble,Xia2004bubble,Deng2012}.






Monolayer graphene has emerged as a versatile platform to study the QHE, showing many of the same phenomenon that for a long time were limited to very high mobility GaAs heterostructures, while also  introducing new opportunities for manipulating these phases owing to the unique combination of a non-trivial $\pi$ Berry phase, four-fold degeneracy arising from the spin and valley iso-spin degrees of freedom, and the ability to fabricate devices in a wide variety of architectures ~\cite{Novoselov2005,Yuanbo2005,Dean2011,Young2012,Feldman2012,Amet2015,Zibrov2017,Hunt.17,Li.17b}. Recent improvements in device fabrication designed to eliminate impurity scattering in sample bulk, such as use of boron-nitride as an improved substrate dielectric~\cite{Dean.10} and fully encapsulated geometries ~\cite{Lei.13,Zibrov2017,Li.17b} have enabled observation of some of the most fragile ground states in the QHE regime ~\cite{Klitzing1980,Cui_FQHE,Du:2009,Dean2011} including the even denominator fractional quantum Hall effect (FQHE) state ~\cite{Zibrov2017,Li.17b,Ki:2014} as well as various electron solid phases ~\cite{Chen2018RIQHE,Smet2018even}. Despite these advancements, the resolution in transport measurement in conventional Hall bar geometries is often overshadowed by measurements that probe the bulk compressibility ~\cite{Feldman2012,Zibrov2017, Hunt.17}. This result is puzzling as it suggests that, contrary to conventional expectation, a well developed bulk gap alone is not a sufficient condition to guarantee well resolved transport measurement of the corresponding edge modes.   


In this work we investigate a less explored aspect of QHE by studying the bulk property of graphene heterostructure using a Corbino geometry\cite{Gervais2015Corbino,Yan2010corbino,Zhao2012corbino,Peters2014corbino,Geim2017corbino,Kumar2018corbino}. We demonstrate a novel fabrication method that allows us to realize concentric contacts in a dual-gated geometry. The successful fabrication of high quality graphene Corbino discs allows us to resolve FQHE states over larger filling fractions and to lower magnetic fields than previously demonstrated in transport measurement of conventional hall bar geometries. Using this technique we identify apparent phase transitions in the FQHE sequence providing new insight about their ground state order in both the lowest and first excited Landau levels (LLs), and demonstrate features consistent with various electron solid phases in higher LLs. Our capability to detect QHE signatures  with higher resolution using the Corbino geometry, where bulk response dominates, compared to Hall bar geometries, where edge transport dominates, suggests that details of the sample edge play a significant role in the Hall bar response.  This result has implications for all transport measurements of 2D topological systems, and suggests our understanding of how to probe edge modes in 2D materials may need to be revisited.

\begin{figure*}
	\includegraphics[width=1\linewidth]{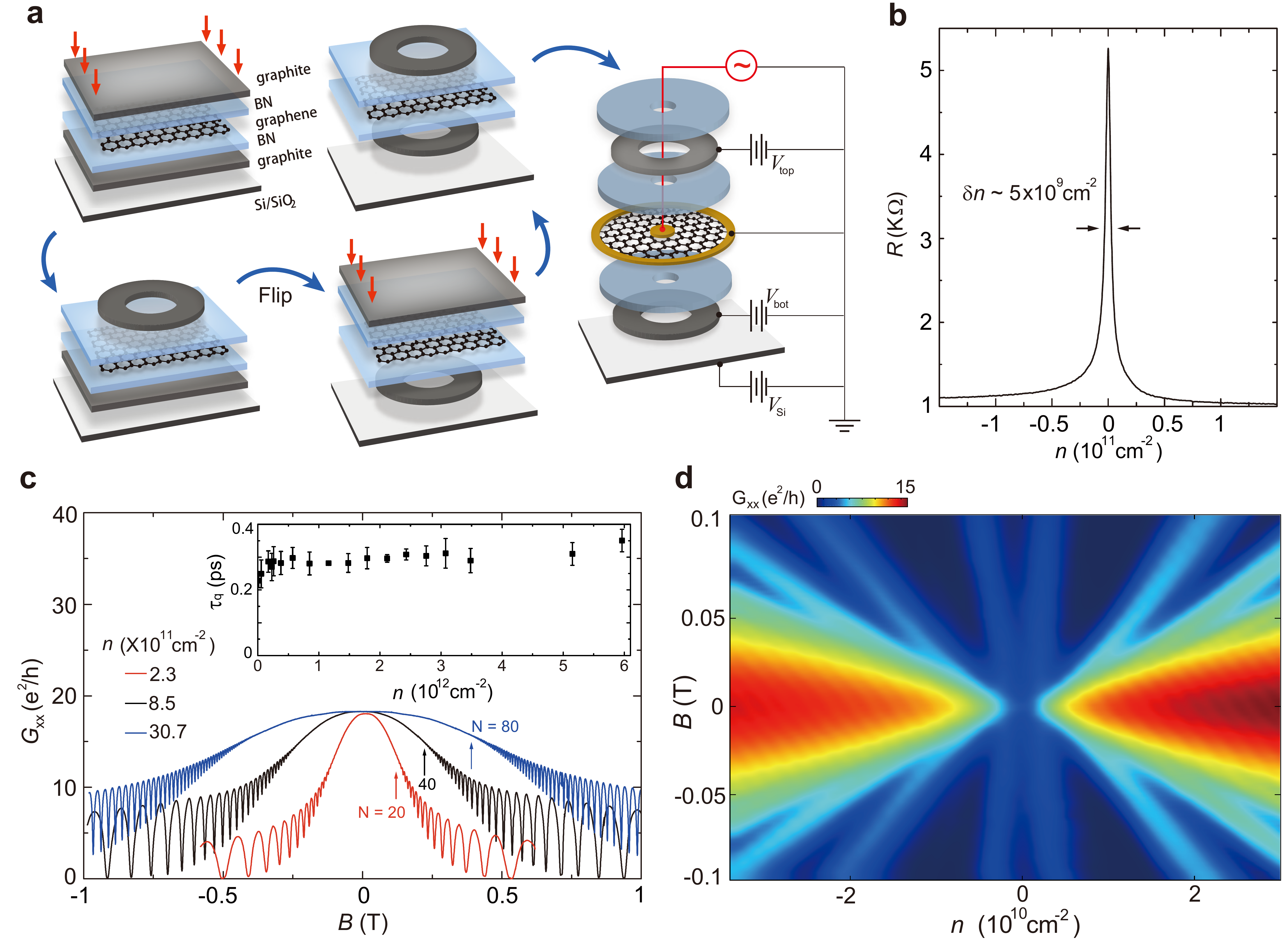}
	\caption{\label{fig1v7}{\bf{Device fabrication and low field characterization.}} (a) Fabrication process of a graphene corbino device with dual-graphite gates.  An hBN-encapsulated heterostructure including top and bottom graphite gates is assembled using a dry transfer method (upper left panel) ~\cite{Lei.13}. The first graphite gate is shaped using plasma etching (red arrows).  The heterostructure is then flipped and the second gate is etched to be aligned to the first and covered with a final hBN layer. The gate electrodes are designed to be slightly smaller than the graphene, so that graphene channel and gate electrodes can be contacted independently using the edge-contact technique ~\cite{Lei.13}. The device reported here has inner radius 1.8~$\mu$m,  outer radius  8.8~$\mu$m and top and bottom BN thicknesses of 57~$n$m and 46~$n$m, respectively.  (b) Bulk resistance as a function of carrier density at $\textit{T} = 2 $ K and $\textit{B} = 0 $ T.   (c) 2-terminal bulk conductance $G_{xx}$ as a function of $B$-field measured at different charge carrier densities $n$, at $\textit{T} = 300 $ mK showing SdH oscillations. Inset, the quantum lifetime,  extracted from SdH oscillations, is plotted against carrier density, $n$. (d) Low field bulk conductance, $G_{xx}$ plotted versus $n$ and $B$-field at $\textit{T} = 300 $ mK.}
\end{figure*}  

\begin{figure*}
	\includegraphics[width=1\linewidth]{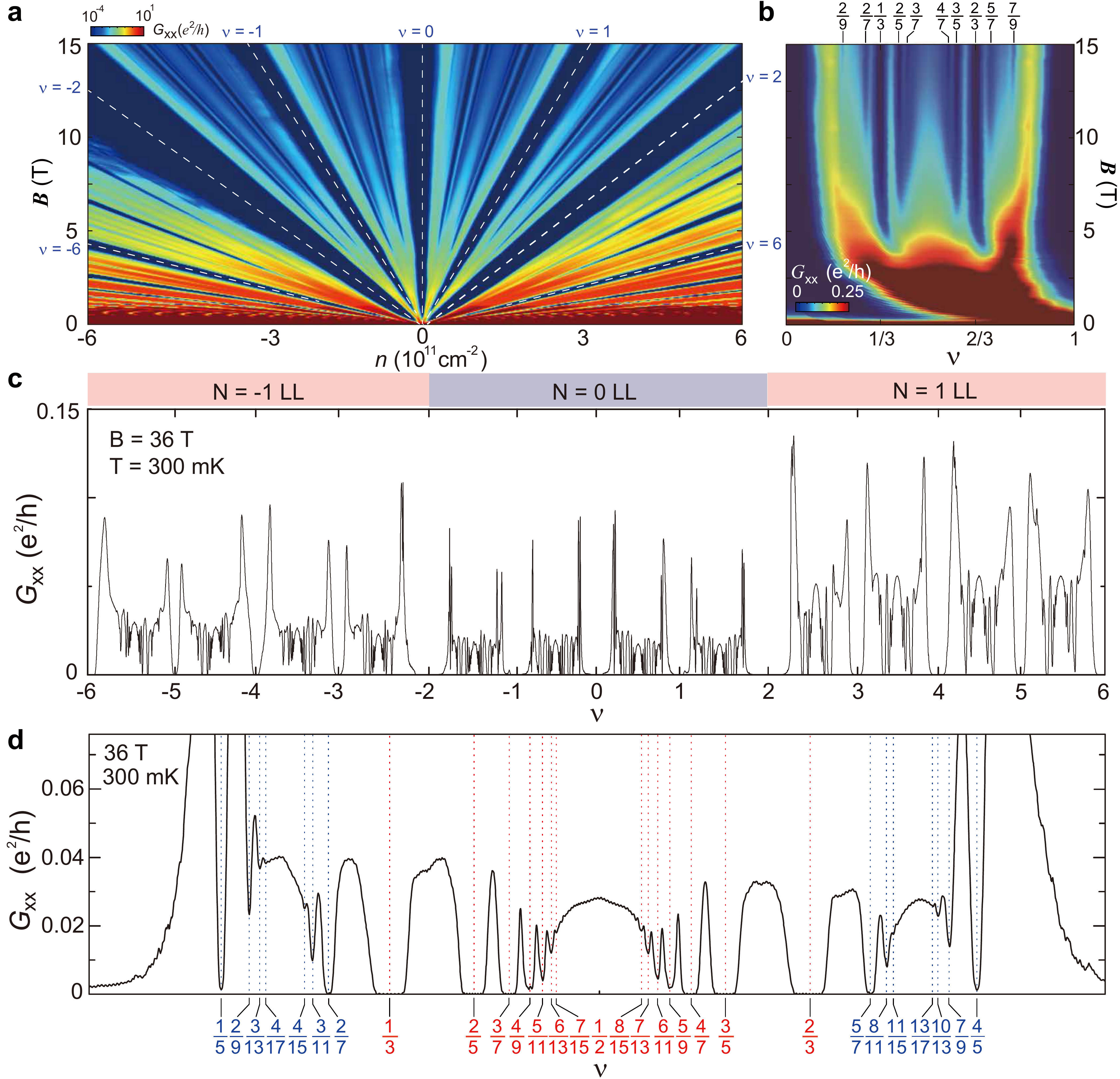}
	\caption{\label{fig2v14} {\bf{Fractional quantum Hall effect.}} (a) Bulk conductance $G_{xx}$, as a function of filling fraction, $\nu$, and $B$ field. (b) $G_{xx}$, versus $\nu$ and $B$ at $T=0.3$ K for $0 < \nu <1$. (c) $G_{xx}$ as a function of filling fraction for the $N=0$ and $N=1$ LL, $-6 < \nu < 6$. (d) High resolution view of one integer branch of the $N=0$ LL at $B = 36$ T and $T=0.3$ K. Red lines and blue lines identify FQHE states belong to the 2-flux and 4-flux composite fermion sequences, respectively.}
\end{figure*}

Fig.~1a illustrates the device fabrication process.  The heterostructure is assembled using the previously described dry transfer technique ~\cite{Lei.13} to ensure clean interfaces between component materials and  includes both top and bottom graphite gates to screen remote impurities and maximize channel mobility ~\cite{Zibrov2017,Li.17b}. The challenge of making  electrical contact to the inner and outer edges of the Corbino geometry is addressed by using a process we refer to as a flip-stack technique (Fig.~1a) (see supplementary material for more details). After the heterostructure is fully assembled the exposed graphite gate is etched into an annulus using standard lithography,  the structure is then flipped over and the second graphite gate is etched so as to be  aligned to the first. The entire structure is then covered with an additional BN layer and a final lithography step is used to realize edge contacts~\cite{Lei.13} to the inner and outer rings of the graphene channel as well as the two graphite gates.  In the final device structure the aligned graphite gates define the carrier density in the active region of the Corbino geometry whereas the densities in the contact regions are tuned by biasing the Si gate. 

Fig. 1b shows resistance versus channel density acquired at $T\sim2$~K, and $B=0$~T. The width of the CNP resistance peak provides an estimate of the charge inhomogeneity ~\cite{Lei.13}, and is found to be $~ 6\times10^{9}$~cm$^{-2}$ (Fig. 1b). This is an order of magnitude lower than previously reported in graphene devices without graphite gates~\cite{Lei.13} but similar to what we measure in  Hall bar devices that include both top and bottom graphite gates (see supplementary material).

\begin{figure*}
	\includegraphics[width=1\linewidth]{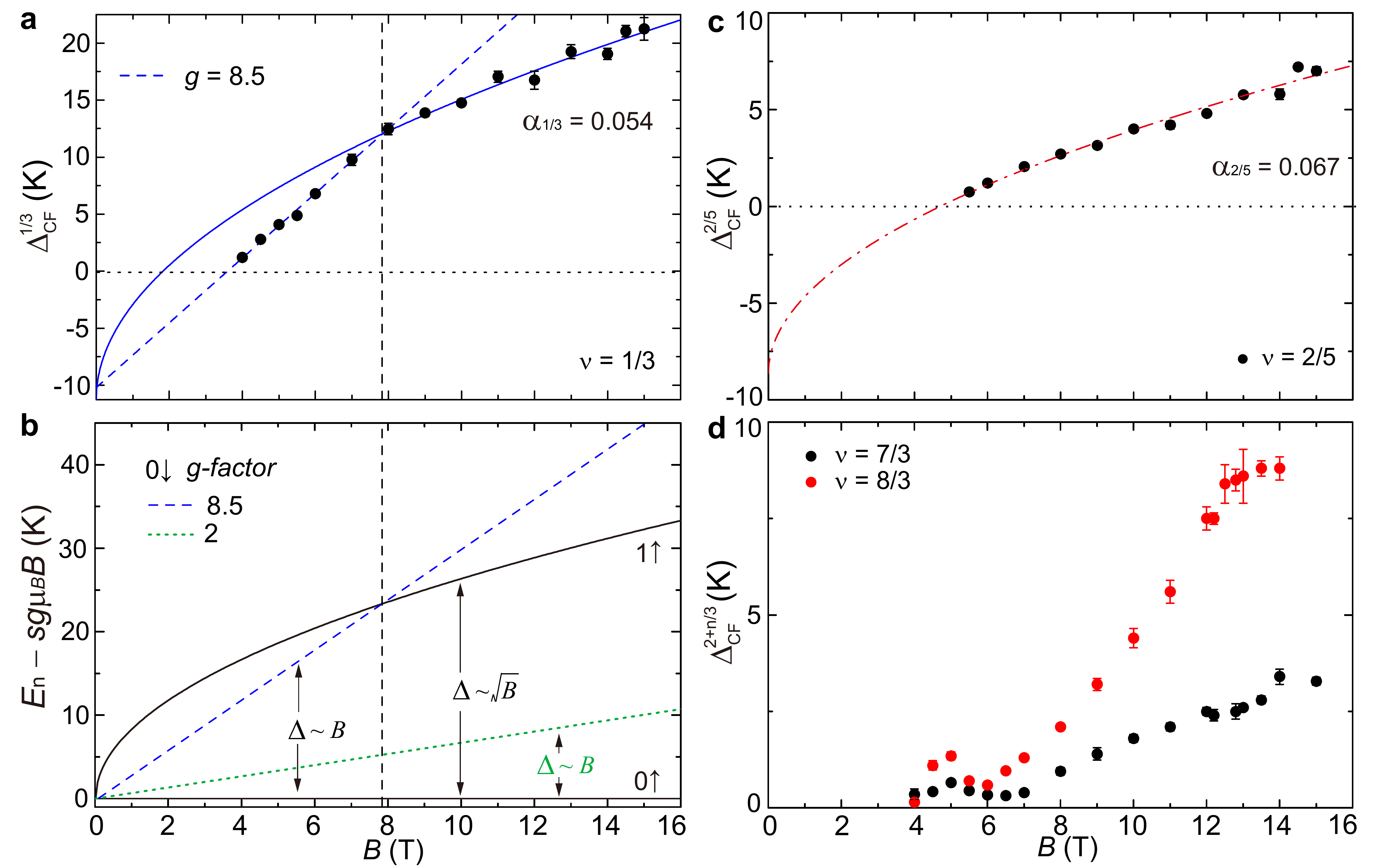}
	\caption{\label{fig3} {\bf{Activation energy gap of FQHE states versus $B$-field.}} (a) Activation energy gap of the $\nu = 1/3$ FQHE state. The blue solid curve is a $\sqrt{B}$ fit to data for $B > 8$ T. The blue dashed line is a linear fit to $B  < 8$ T. (b) Energy level diagram.  The levels are labeled for CF cyclotron orbits with different spin polarization and $g-$factor as a function of $B$ field.   The two lowest CF cyclotron levels are labeled by their CF orbital index, $N=0$ and $1$, and spin polarizations,  $\uparrow$ and $\downarrow$ . (c) Activation energy gap of the $\nu = 2/5$ FQHE state. The red dash-dotted line is a $\sqrt{B}$ fit to the data. (d) Activation energy gap of the $\nu = 7/3$ and $8/3$ FQHE state in the $N=1$ LL.}
\end{figure*}

Fig. 1c shows the low magnetic field Shubnikov de Haas (SdH) oscillations for three representative densities. Extraction of the quantum scattering time $\tau_q$ from the corresponding dingle plots (see supplementary material) shows a relatively density independent value of $\tau_{q}\sim0.3$~ps, except at very low densities (Fig.~1c inset) where it falls off. This value of $\tau_{q}$ is among the largest values reported for graphene, further confirming the low bulk disorder in our sample. An independent estimate of the quantum lifetime can be made by assuming that the SdH onsets when the field-dependent cyclotron gap, $\Delta_c$,  exceeds the LL disorder broadening, $\Gamma$, where $\Gamma=\hbar/{2\tau_{q}}$ and, for graphene, $\Delta_c\sim~400\sqrt{B}\sqrt{N}$, where $B$ is the magnetic field and $N$ is the LL orbital index. This estimate also gives a mostly density-independent value of $\Gamma\sim15$~K which agrees well with the saturated value of 12~K obtained from the measured $\tau_q$ (a full density dependent comparison is shown in the supplementary material).

Fig. 1d shows a Landau fan diagram in the low density and low magnetic field regime. In a Hall bar geometry, two and four terminal measurements probe the resistance associated with both the bulk and dissipationless edge modes. By contrast, a Corbino geometry, where there are no physical edges, probes only the bulk conductance.  In this case fully developed, incompressible,  QHE ground states  manifest as zero conductance (infinite resistance). The fan diagram in Fig.~1d shows well developed QHE states (nearly zero conductance) at filling fraction $\nu=\pm 2$ emerging at fields less than $B \sim 50$ mT. Fig.~2a shows a similar Landau fan diagram but measured over a larger density and field range.  Several distinguishing features are evident: the plot shows excellent ambipolar response with both electron and hole features equally resolved; the symmetry broken IQHE emerge at less than $B=1$ T; and the FQHE is resolvable by $B=5$ T (Fig. 2b). This quality of QHE transport has been difficult to achieve in Hall bar geometries, even when the sample disorder is similar as measured by zero field transport and SdH characteristics (see supplementary material).  

The origin of the improved resolution obtained in our Corbino geometry may be two-fold. First, the Hall bar measurement requires good electrical contact~\cite{Klitzing2011}, since the leads should be well equilibrated to the edge modes in order to measure zero longitudinal resistance and accurate Hall plateau.  This is a less stringent requirement in the Corbino geometry where QHE ground state appears as an insulating feature in bulk conductance, even for highly resistive contacts.  Second, transport measurement of the edge state may be complicated by details of the potential profile near the graphene boundary ~\cite{Cui2016,Geim2017corbino},  edge disorder~\cite{Neto2006}  and edge mode reconstruction~\cite{Sabo2017}. 


The improved performance of the Corbino geometry allows us to resolve the FQHE states in graphene to an unprecedented degree, particularly in the high field/low density limit.  In Fig. 2c the bulk conductance, $G_{xx}$, is plotted versus density at $B=36$~T. In both  the $N=0$ LL and $N=1$ LLs, standard composite fermion (CF) sequences are observed ~\cite{Jain.89}, including both even and odd numerator FQHE states, indicating that all symmetries have been lifted ~\cite{Dean.10,Feldman.13}. Fig. 2d shows an expanded view in the $N=0$ LL between $\nu=0$ and $\nu=1$.  Two-flux CF states (centered around $\nu=1/2$) and four-flux CF states (centered around $\nu=1/4$) up to denominator 15 are observed.  We note that based on the depth of the conductance minima, the overall hierarchy appears remarkably electron-hole symmetric, further indicating that all symmetries are lifted within the CF levels (this is confirmed by activation gap measurements, which show a similar hierarchy, see supplementary material). A different symmetry is observed  in the $N=1$ LL, suggesting that the spin and valley degeneracy is only partially lifted, and an approximate SU(2) or SU(4) symmetry is preserved for the composite fermion ground states.

The persistence of the strongest FQHE states to low magnetic fields allows us to measure how their gaps evolve over a wide range of $B$.  Fig.~3a shows a plot of the activation energy gap, $\Delta$, versus $B$,  for the $\nu=1/3$ state.  A clear kink in the trend is observed at  $\textit{B} \sim 8$~T below which the gap is best fit by a linear  $B$ dependence (blue dashed line) and above which the gap transitions to  a $\sqrt{B}$ dependence (blue solid curve). Notably, both the linear and square-root fits extrapolate to $\Delta \sim -10$ K at $B = 0$, similar to the value of disorder broadening estimated from the SdH behavior (Fig.~1c), providing a self consistency validation of the fits.

The transition in the $B$ dependence of the gap resembles similar behavior of the $1/3$ FQHE state in GaAs quantum wells, which was interpreted in the context of CF Landau levels with spin degrees of freedom~\cite{Dethlefsen2006, Haug2004}. In the CF picture, the effective cyclotron gap that separates spin-degenerate CF LLs results from Coulomb interaction and is given by ~\cite{Haug2004}, $\Delta_{CF}^{cyclotron} = \dfrac{{\hbar}e\textit{B}^{\ast}}{m^{\ast}}$ where $B^{*}=B-B_{\nu=1/2}$ is the effective magnetic field for CFs and  $m^{\ast} = \alpha m_e \sqrt{B}$ is the CF mass, $m_{e}$ is the free electron mass and $\alpha$ depends on details of the quantum well.  Allowing for spin degree of freedom, the CF LLs can split into spin branches, separated by the Zeeman energy $E_{CF}^{Zeeman}= \frac{1}{2}\mu_{B}gB$, where $\mu_{B}$ is the Bohr magneton and $g$ is the Lande $g$-factor. The transition results from a CF LL crossing when the CF Zeeman energy (linear in B), exceeds the CF cyclotron energy (square root in B), as illustrated in Fig. 3b. This model well fits our data in the Lowest LL.  If we assume that the linear trend correlates to a real spin gap, the slope gives an estimate for the $g$-factor of $8.5$.  This is approximately 4 times larger than the bare electron $g$-factor ($g=2$),  and is indicative of strong exchange interaction and the existence of skyrmion spin textures for composite fermions ~\cite{Young2014}.  In this picture we imagine that the valley degrees of freedom is frozen out ~\cite{Feldman.13} such that the square root region corresponds to the CF cyclotron gap.  
Fitting the above expression to this region gives a CF mass term of $\alpha=0.054\pm 0.004$.  Including the projected disorder broadening of $\sim10$~K, this gives a measure of the intrinsic gap to be $\Delta_{1/3} = (8.3 \pm 0.6)\sqrt{B}$~K, or  $(0.084 \pm 0.004)$ $e^{2}/\epsilon l_{B}$ in Coulomb energy units, where we use $\epsilon=6.6$ for BN-encapsulated graphene~\cite{Hunt.17}.  We  note that this result is remarkably close to the theoretical value of 0.1 $e^{2}/\epsilon l_{B}$ calculated by exact diagonalization~\cite{Morf2002} without including any  additional corrections ~\cite{Balram2015} (see supplementary material for detailed comparison).

\begin{figure}
	\includegraphics[width=1\linewidth]{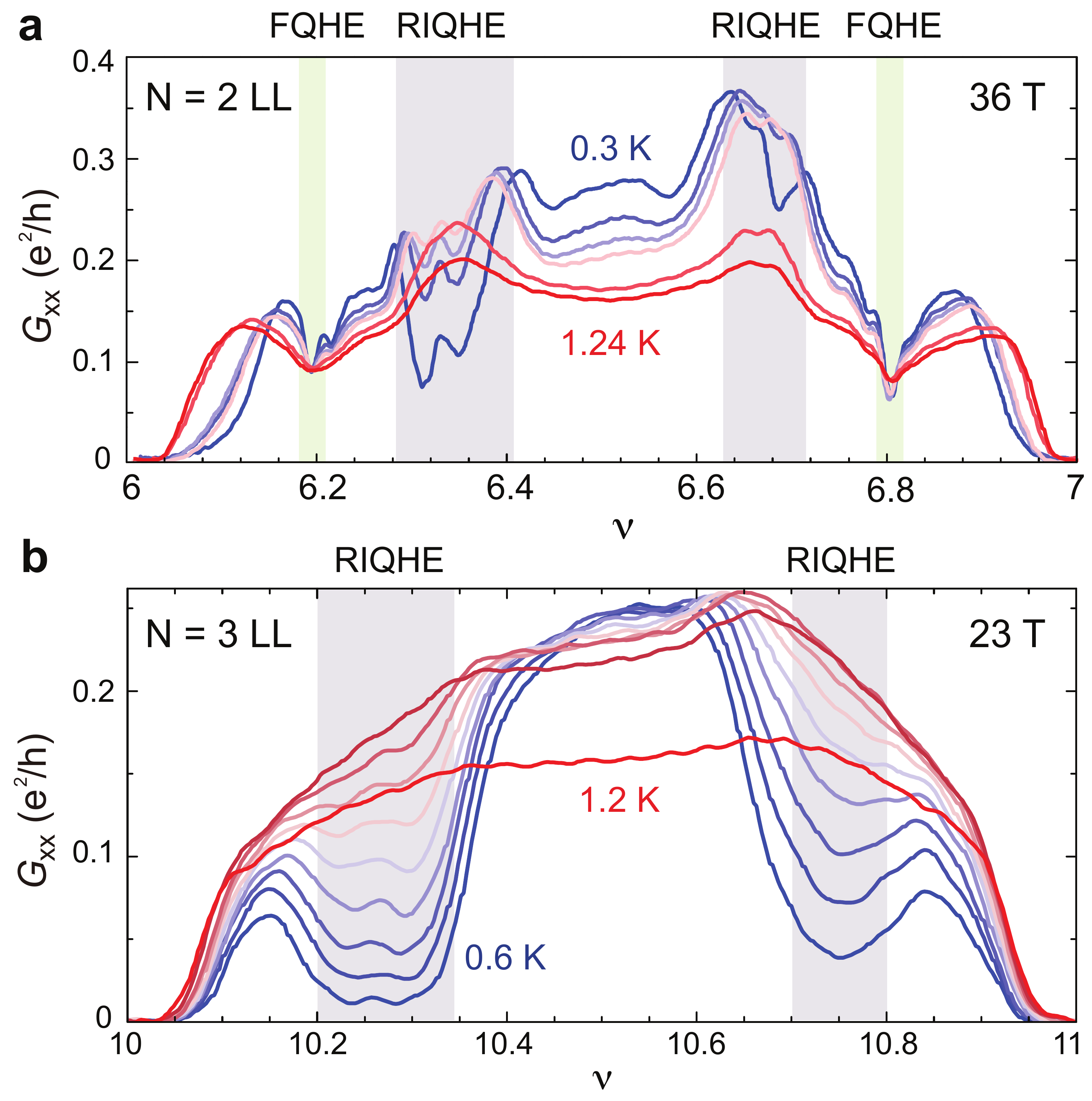}
	\caption{\label{fig4} {\bf{RIQHE in N = 2 and 3 LL.}} Bulk conductance as a function of filling fraction measured at different temperature in (a) N = 2 LL and  (b) N = 3 LL respectively. }
\end{figure} 

Fig. 3c shows the $B$ dependence of the $\nu=2/5$ gap.  In this case the gap follows a $\sqrt{B}$-dependence over the entire accessible field range, projecting to a $\sim -10$ K at $B = 0$.  The disorder broadening is consistent with measurement of the $1/3$ gap and SdH analysis. The square root dependence is qualitatively consistent with same CF picture as above in which the $\nu=2/5$ represents a cyclotron gap of CF LLs 
 In this view however, the lack of a transition is surprising (we would expect the CF cyclotron gap to show evidence of the same CF LL crossing that gives rise to the kink in the 1/3 gap, see supplementary material), and may suggest that the exchange interaction for CFs is highly sensitive to composite fermion filling fraction\cite{Hunt.17}.
 
In the $N = 1$ LL, a phase transition is observed for $\nu = 8/3$ where the energy gap vanishes at $B \sim 6$ T and then reemerges at higher field (Fig. 3d). Similar transition with vanishing energy gap was also observed in local electron compressibility measurement of suspended graphene  ~\cite{Feldman.13}. Such behavior cannot be understood within the schematic energy diagram shown in Fig.~3b and is likely related to transition between different iso-spin polarizations.  A complete understanding of this phase transition will require definitive identification of the iso-spin order associated with the CF states ~\cite{Young2012}. 



Fig.~4 plots bulk conductance measured at higher Landau levels. In the $N=2$ LL, we observed features corresponding to 4-flux CF ground states at $\nu=6+\frac{1}{5}$ and $6+\frac{4}{5}$ and electron solid states at $\nu=6+\frac{1}{3}$ and $6+\frac{2}{3}$. The electron solid state is characterized by the non-monotonic temperature dependence in the bulk conductance, with a peak at the melting transition $T_c$ which diminishes to zero at low temperature ~\cite{Eisenstein2002bubble,Deng2012,Chen2018RIQHE}. 
In the $N=3$ LL, bulk conductance displays a broad minimum around $\nu=10+\frac{1}{4}$ and $10+\frac{3}{4}$ as shown in Fig.~4b, where the temperature evolution resembles the bubble phase of $N=3$ LL observed in GaAs samples with Corbino geometry ~\cite{Gervais2015Corbino}. The deep conductance minima observed in the Corbino geometry and the high transition temperature of $\sim 1.1$ K are both indicative of a robust electron solid state, which is qualitatively similar to recent measurement in MLG samples with a Hall bar geometry ~\cite{Smet2018even}. Interestingly, the bulk conductance reveals no obvious feature at half filling down to $T=0.3$ K at $B = 25$ T (see supplementary material), which is in contrast to the even-denominator state recently reported in MLG samples with Hall bar geometry ~\cite{Smet2018even}. Given the high resolution and large energy gap of correlated states observed in Corbino geometry, a potential electron liquid phase such as the Pfaffian is expected to show up as a sharp minimum in bulk conductance.

In summary we have established a process of realizing very high quality Corbino devices in a dual-gated geometry. The ability to directly probe bulk conductance in the QHE regime, independently of the edge states, provides new access to various electron liquid and solid states in graphene beyond what has previously been possible in transport studies. Additionally, the superior quality compared to similarly constructed Hall bar devices suggests that transport measurement in the conventional Hall geometry is limited by difficulties related to probing the edge channels but not bulk disorder.  This might be due to details of the edge mode structure\cite{Sabo2017} or difficulties in designing contacts that well equilibrate to the edge channels~\cite{Klitzing2011}.


\section*{acknowledgments}
We thank Andrea Young for helpful discussions and sharing unpublished results.  This work was supported by the ARO under MURI W911NF-17-1-0323. A portion of this work was performed at the National High Magnetic Field Laboratory, which is supported by National Science Foundation Cooperative Agreement No. DMR-1644779 and the State of Florida.  CRD acknowledges partial support  by the David and Lucille Packard Foundation.

\section*{Competing financial interests}
The authors declare no competing financial interests.

\end{document}